\documentclass[twocolumn,amsmath,amssymb,superscriptaddress,prb]{revtex4}

\usepackage{graphicx}
\usepackage{times}
\usepackage{amsmath,amssymb,amsthm,amsfonts}
\usepackage{bbold}

\renewcommand{\vec}[1]{\boldsymbol{#1}}
\newcommand{\beq}{\begin{equation}}
\newcommand{\eeq}{\end{equation}}
\newcommand{\beqa}{\begin{eqnarray}}
\newcommand{\eeqa}{\end{eqnarray}}
\newcommand{\e}{\mathrm{e}}

\newcommand{\ket}[1]{\left| #1 \right\rangle}
\newcommand{\bra}[1]{\left\langle #1 \right|}

\newcommand{\ketbra}[2]{\left|#1\right\rangle\hskip-1mm\left\langle #2\right|}

\newcommand{\plus}{\Psi}
\newcommand{\moins}{\Phi}

\begin{document}

\title{Noise effects and tomography of remote entangled spins in quantum dots}
\author{Asier Pineiro-Orioli}
\affiliation{Optics Section, Blackett Laboratory, Imperial College London, London SW7 2AZ, UK}
\author{Dara P. S. McCutcheon}
\affiliation{Optics Section, Blackett Laboratory, Imperial College London, London SW7 2AZ, UK}
\affiliation{Departamento de F\'isica, FCEyN, UBA and IFIBA, Conicet, Pabell\'on 1, Ciudad Universitaria, 1428 Buenos Aires, Argentina}
\author{Terry Rudolph}
\affiliation{Optics Section, Blackett Laboratory, Imperial College London, London SW7 2AZ, UK}

\date{\today}

\begin{abstract}

We investigate how decoherence affects the entanglement established between two quantum dots in micro cavities, 
and propose a tomographic scheme able to measure the entangled state. The scheme we consider, establishes the  
entanglement via the exchange and measurement of a photon.
%We investigate decoherent effects on, and propose a tomographic procedure to measure the 
%entangled state of two quantum dots in micro cavities, 
%established via the exchange and subsequent measurement of a photon. 
Making the realistic assumption of noise dominated by pure-dephasing processes, 
we find that 
the photon must be exchanged and measured on timescales shorter than the quantum dots' characteristic dephasing time for 
appreciable levels of entanglement to be achieved.
The tomographic scheme is able to reconstruct the full density matrix of the quantum dots, 
and requires only single spin rotations and the injection of an additional photon. 
Remarkably, we find that the additional photon need not be exchanged and measured on a 
timescale shorter than the dephasing time for accurate tomography, 
and also allows many to be used in order to increase the measurement signal.

\end{abstract}

\maketitle

\section{Introduction}

Entanglement is a resource in the field of quantum information and plays a particularly important role in quantum communication and 
quantum networks.~\cite{kimble08,n+c} 
As such, there has been a huge amount of research into schemes which are able to effectively generate entanglement between spatially separated systems. 
Examples include using a direct coupling between the two systems,~\cite{imamoglu99,kraus01,Calarco03} through projective measurements,~\cite{browne03,cabillo99,barret05,bose99,plenio99} 
and even via correlated dissipative processes.~\cite{benatti03,mccutcheon09,plenio02}

One such implementation is the generation of remote spin-spin entanglement shared between semiconductor quantum dots (QDs) in microcavities via the 
exchange and subsequent measurement of a photon.~\cite{bristol_dots,bristol_photons,bristol_tele,young13} 
The spin dependent optical transitions of a charged QD can cause linearly polarised photons 
to undergo giant Faraday rotations, which can be utilised to construct QD-photon entanglement. Allowing a photon entangled 
with one such QD to interact with another QD, and subsequent measurement of the photon can cause the two QDs to become entangled. 
This scheme is promising since the subjects of the entanglement - the 
QDs - constitute good candidates for the storage and manipulation of quantum information.~\cite{loss98,Calarco03} The photon which transmits the entanglement, on 
the other-hand, is an ideal candidate to transmit quantum information owing to its intrinsically long coherence time.~\cite{cirac97,knill01,kok07} 
We note that the understanding and ability to control QD-cavity systems is advancing considerably, with resonance fluorescence,~\cite{vamivakas09,ulrich11_short,mccutcheon13} 
quantum dot induced phase shifts,~\cite{young11} 
and more recently QD-photon entanglement~\cite{greve12,gao12} having been measured and characterised. 

In the ideal case of the entangling procedure mentioned above, maximally entangled states of the two QDs can be achieved. 
One source of errors which may cause a deviation from this ideal scenario 
is the unavoidable coupling of the QDs to their solid-state environments. In typical In(Ga)As QDs, the coupling of an excess electron 
to the nuclear spins in the QD semiconductor via the hyper-fine interaction is thought to 
dominate over other sources of decoherence, such as coupling to bulk phonons via spin-orbit interactions.~\cite{t1_estimate2,golovach04,khaetskii01} 
This can cause a loss of coherence of the electron, which will necessarily impair the entangling procedure described 
above. Thus, it is important to establish what limitations noise is likely to put on the entanglement which can be achieved, and 
to determine how any possible entanglement may be experimentally verified. 

In this work we investigate the effects of decoherence on the level and type of entanglement which can be obtained. In addition, we propose 
a tomographic scheme which would allow for the complete reconstruction of the QD-QD density matrix after the entangling procedure has been performed. 
Our tomographic scheme relies only on single spin rotations of the QDs, and the injection of an additional photon. For the pure-dephasing noise considered, 
we find that the tomography is unaffected by the time taken for the second photon to be exchanged and measured. As such, the 
tomographic procedure reveals the true state of the QDs after entanglement has been established. In addition, we show that it 
is possible to use many photons for the tomography in order to boost the measurement signal.

\section{Summary of entanglement scheme}

The entanglement scheme first proposed in Ref.~[\onlinecite{bristol_dots}] relies on the spin dependent optical transition 
from the ground state of a singly charged QD to an excited trion state ($X^{-}$). The trion state consists of a pair of electrons each having angular momentum 
projection $\pm \hbar/2$, bound to a heavy hole which has angular momentum projection $\pm 3\hbar/2$.~\cite{lovett05,pazy03} There are 
therefore two excited states, $\ket{\uparrow \downarrow, \Uparrow}$ and $\ket{\uparrow \downarrow, \Downarrow}$, having angular momentum projection 
$+3\hbar/2$ and $-3\hbar/2$ respectively. Simple application of the Pauli exclusion principle and conservation of angular momentum, 
reveals that an incident photon having left-circular polarisation, $\ket{L}$, and angular momentum $+\hbar$, can 
excite the $\ket{\uparrow}\to \ket{\uparrow \downarrow, \Uparrow}$ transition, but does nothing to $\ket{\downarrow}$. Similarly, 
a right circularly polarised photon $\ket{R}$ can excite the $\ket{\downarrow}\to \ket{\uparrow \downarrow, \Downarrow}$ transition, but otherwise does nothing.
Placing such a QD inside a cavity causes left and right circularly polarised photons to acquire different phase shifts, depending on 
the spin orientation of the electron in the QD.~\cite{bristol_dots,bristol_photons,bristol_tele}

Following Ref.~[\onlinecite{bristol_dots}], we ignore the side leakage of the cavity $\kappa_s$ (or $\kappa_s<\kappa$), 
where $\kappa/2$ is the cavity field decay rate into the input/output modes, and work in a regime with 
$|\omega-\omega_c|\ll g$ and $g>(\kappa,\gamma)$, where 
$\omega$ and $\omega_c$ are the frequencies of external field and cavity mode, $g$ is the coupling strength between the 
trion $X^-$ and the cavity mode and $\gamma/2$ is the $X^-$ dipole decay rate. 
In doing so, a linearly polarised photon undergoes a giant Faraday rotation with near unity reflectance. 
The whole process can be expressed as the unitary 
\begin{equation}
U_i(\varphi)=\exp[i\varphi(|L\rangle\langle L|\otimes\ketbra{\uparrow}{\uparrow}_i+|R\rangle\langle R|\otimes\ketbra{\downarrow}{\downarrow}_i)],
\label{eq:unitary}
\end{equation}
where $\varphi$ is the difference in phase shifts experienced by left and right circular polarisations, and the index indicates interaction 
of the photon with the ith QD. 
By adjusting $\omega-\omega_c\approx\pm\kappa/2$, giant Faraday rotations of $\varphi=\pm\frac{\pi}{2}$ can be achieved.~\cite{bristol_dots}
From a quantum information perspective Eq.~({\ref{eq:unitary}}) constitutes an entangling gate between the QD and the incident photon. 
By allowing a photon to interact with one such QD in a cavity followed by another, an entangled state of 
the photon and both QDs is produced. To see this, we take a vertically polarised initial state of the photon, 
$|\phi_{\mathrm{ph}}\rangle=\ket{V}=(1/\sqrt{2})(|R\rangle+|L\rangle)$, while that of QDs as 
$|\phi_{i}\rangle=\alpha_i|{\uparrow}\rangle_i+\beta_i|{\downarrow}\rangle_i$ for $i=1,2$.
Thus, the initial state of the whole system is 
$|\phi_{\mathrm{tot}}\rangle=|\phi_{\rm{ph}}\rangle\otimes|\phi_{1}\rangle\otimes|\phi_{2}\rangle$, 
and following the interaction of the photon with both QD-cavity systems we have 
$\ket{\phi_f}=U_{2}(\textstyle{\frac{\pi}{2}})U_{1}(\textstyle{\frac{\pi}{2}})|\phi_{\mathrm{tot}}\rangle$ giving 
\begin{align}
\ket{\phi_f}=\ket{H}&[\alpha_1\alpha_2|{\uparrow}\rangle_1|{\uparrow}\rangle_2
-\beta_1\beta_2|{\downarrow}\rangle_1|{\downarrow}\rangle_2]\nonumber\\
+i\ket{V}&[\alpha_1\beta_2|{\uparrow}\rangle_1|{\downarrow}\rangle_2+ \alpha_2\beta_1|{\downarrow}\rangle_1|{\uparrow}\rangle_2],
\label{eq:sys_evolution}
\end{align}
where $\ket{H}=(1/\sqrt{2})(\ket{R}-\ket{L})$ is a horizontally polarised photon. 
Thus, when measuring the photon in the linearly polarised basis, $\{\ket{V},\ket{H}\}$ the state of the QDs collapses to the (normalised) states
\begin{align}
\label{eq:end_state_moins}
|\Phi\rangle &=\sqrt{2}\Big(\alpha_1\alpha_2|{\uparrow}\rangle_1|{\uparrow}\rangle_2- \beta_1\beta_2|{\downarrow}\rangle_1|{\downarrow}\rangle_2\Big) \ \ \text{for}\ \ket{H}\ \text{or}\\
\label{eq:end_state_plus}
|\Psi\rangle &=\sqrt{2}\Big(\alpha_1\beta_2|{\uparrow}\rangle_1|{\downarrow}\rangle_2+ \alpha_2\beta_1|{\downarrow}\rangle_1|{\uparrow}\rangle_2\Big) \ \ \text{for}\ \ket{V}.
\end{align}
Upon setting $\alpha_{1,2}$ and $\beta_{1,2}$ to $1/\sqrt{2}$, we find 
$\ket{\Phi}\to\ket{\phi^-}=(1/\sqrt{2})(\ket{\uparrow\uparrow}-\ket{\downarrow\downarrow})$ while 
$\ket{\Psi}\to\ket{\psi^+}=(1/\sqrt{2})(\ket{\uparrow\downarrow}+\ket{\downarrow\uparrow})$, which are maximally entangled Bell states. 
We call Eq.~(\ref{eq:end_state_moins}) the `$\moins$' outcome and Eq.~(\ref{eq:end_state_plus}) the `$\plus$' outcome, alluding 
to symmetry present in the corresponding state of the QDs in this idealised case.

\section{Noise effects on the entanglement generation}

Matters relating to the imperfect implementation of the operation described by Eq.~({\ref{eq:unitary}}) have been discussed in Ref.~[\onlinecite{bristol_dots}]. 
In this section, our aim is to investigate the non-unitary dynamical evolution of the QDs during the entangling procedure. 
We therefore assume that the QD-photon interaction happens instantaneously, and consider three time intervals during which 
the QDs undergo decoherent processes: 
$t_1$ labels the time from the release of the photon until it reaches the first QD; $t_2$ is the time taken for 
the photon to travel from the first QD to the second, and $t_3$ the time from the second interaction until the photon is measured. 
During each of these times the combined photon-QDs system is assumed to undergo non-unitary evolution generated by 
a master equation of Lindblad form~\cite{b+p} (we set $\hbar=1$):
\begin{equation}
\dot{\rho}=-i[H,\rho]+\sum_l \left(L_l\rho L_l^\dagger-\frac{1}{2}\{\rho,L_l^\dagger L_l\}\right),
\label{eq:lindblad}
\end{equation}
where $\rho$ is the density matrix of the combined photon-QDs system, $H$ is the free Hamiltonian, and the set $\{L_l\}$ are Lindblad 
operators describing the decoherent processes we wish to consider. We assume the QDs to be under the influence of a magnetic field 
along the $z$-axis such that the Hamiltonian for the QDs system reads 
$H=\frac{\epsilon}{2}(\sigma^z_1+\sigma^z_2)$,
where 
$\sigma^z_i=|{\uparrow}\rangle\langle{\uparrow}|_i-|{\downarrow}\rangle\langle{\downarrow}|_i$ and $\epsilon$ is the field strength. 
We assume the initial state of the entire system is the separable pure state $|\phi_{\mathrm{tot}}\rangle$ given above, 
with the coefficients $\alpha_{1,2}$ and $\beta_{1,2}$ set to $1/\sqrt{2}$. 
Following both photon-QD interactions and 
the three periods of time for which the system evolves according to Eq.~({\ref{eq:lindblad}}), the final state of the entire system is 
\beq
\rho_{\rm{f}}=\e^{\mathcal{L} t_3}\big[U_2\big(\e^{\mathcal{L} t_2}\big[U_1\big(\e^{\mathcal{L}t_1} \ketbra{\phi_{\rm{tot}}}{\phi_{\rm{tot}}}\big)U_1^{\dagger}\big]\big)U_2^{\dagger}\big]
\label{rhof}
\eeq
where the Liouvillian super-operator is defined to satisfy $\dot{\rho}=\mathcal{L}\rho$ in Eq.~({\ref{eq:lindblad}}).
We then measure out the photon and 
analyse the post-measurement ensemble $\{p_k,\rho_k\}$, for $k=\Phi,\Psi$ representing the two measurement outcomes. 
The probability for each outcome is given by $p_k=\mathrm{Tr}[\rho_{\rm{f}} (\pi_k \otimes \openone\otimes\openone)]$, 
where $\pi_{\Phi}=\ketbra{H}{H}$ and $\pi_{\Psi}=\ketbra{V}{V}$, while the post-measurement 
state itself is $\rho_k=\mathrm{Tr}_{\mathrm{ph}}[\rho_{\rm{f}} (\pi_k\otimes \openone\otimes\openone)]/p_k$, where $\openone$ is the identity operator, 
and $\mathrm{Tr}_{\mathrm{ph}}$ denotes a trace over the photon degrees of freedom only.

For electron spins in QDs, experiments have measured relaxation times as long as $T_1\sim \rm{ms}$.~\cite{t1_estimate2,t1_estimate} 
In separate experiments, 
spin coherence decay timescales have been measured to be $T_2\sim\mu\rm{s}$, at best.~\cite{t2_estimate} As such, we expect any 
superposition of $\ket{\uparrow}$ and $\ket{\downarrow}$ - as is required for the QD-entanglement scheme - to be affected 
by pure-dephasing processes on a timescale $T_2$, well before spin relaxation processes become important. We 
therefore consider the pure-dephasing form of Eq.~({\ref{eq:lindblad}}), which is achieved with the pair of 
Lindblad operators $L_i=\sqrt{\Gamma_2/2}\sigma^z_i$ for $i=1,2$, and 
where $\Gamma_2=1/T_2$.

We first remark on two important features. Firstly, the 
measurement probabilities are unaffected by the non-unitary evolution of the QDs; the probability that the post-measurement state 
is $\rho_{\Phi}$ or $\rho_{\Psi}$ remains the same and equal to $1/2$ for all times. This is perhaps to be expected, since pure-depasing does 
not affect the diagonal elements of the QD density matrix when expressed in the basis $\{\ket{\uparrow},\ket{\downarrow}\}$, and 
it is these states through which the QDs are coupled to the photon. 
Put another way, although there is an asymmetry between the dots owing 
to the fact that the photon interacts with one first, this asymmetry cannot be distinguished by pure-dephasing noise and the form of QD-photon 
interaction. Our second remark, and another manifestation of these symmetry arguments, is that the post-measurement 
states depend only on the sum of the three time intervals, $t=t_1+t_2+t_3$. Thus, for pure-dephasing noise considered here, 
there is no particular waiting period which affects the post-measurement states more than the others.

The entanglement fidelity of the two post-measurement states is defined as
 $F_k=\sqrt{\bra{k}_0 \rho_k \ket{k}_0}$, for $k=\Phi,\Psi$, where $\ket{\Phi}_0=\ket{\phi^-}$ and $\ket{\Psi}_0=\ket{\psi^+}$ 
 are the states obtained in the absence of noise.~\cite{n+c} From Eq.~({\ref{rhof}}) we find
\begin{align}
\label{eq:fid_moins}
F_\moins&=\sqrt{\frac{1}{2}\left(1+\e^{-2 t/ T_2}\cos{(2\epsilon t)}\right)},\\
\label{eq:fid_plus}
F_\plus&=\sqrt{\frac{1}{2}\left(1+\e^{-2 t/ T_2}\right)},
\end{align}
and where the cosine factor appearing for $F_\moins$ is present since the magnetic field in the $z$-direction causes oscillations between the entangled states 
$\ket{\phi^-}$ and $\ket{\phi^+}=(1/\sqrt{2})(\ket{\uparrow\uparrow}+\ket{\downarrow\downarrow})$. 
We quantify the entanglement itself using the entanglement of formation,~\cite{wootters98} given by, 
\beq
E(t)=-x(t)\log_2 [x(t)]-(1-x(t))\log_2[1-x(t)]
\eeq
defined in terms of the function $x(t)=(1/2)(1+\sqrt{1-C(t)^2})$ which itself depends on the concurrence $C(t)$, which in the 
present case takes on the simple form $C(t)=\e^{-2 t/T_2}$. 
We find that both the entanglement and the entanglement fidelity decrease in an exponential fashion with 
time.~\cite{mccutcheon09,yu02,solenov07} The non-zero value of fidelity approached reflects that the large $t$ limit of the 
QD system is the maximally mixed state which has non-zero overlap with the target states  $\ket{\phi^-}$ and $\ket{\psi^+}$.
\begin{figure}
\begin{center}
\includegraphics[width=0.35\textwidth]{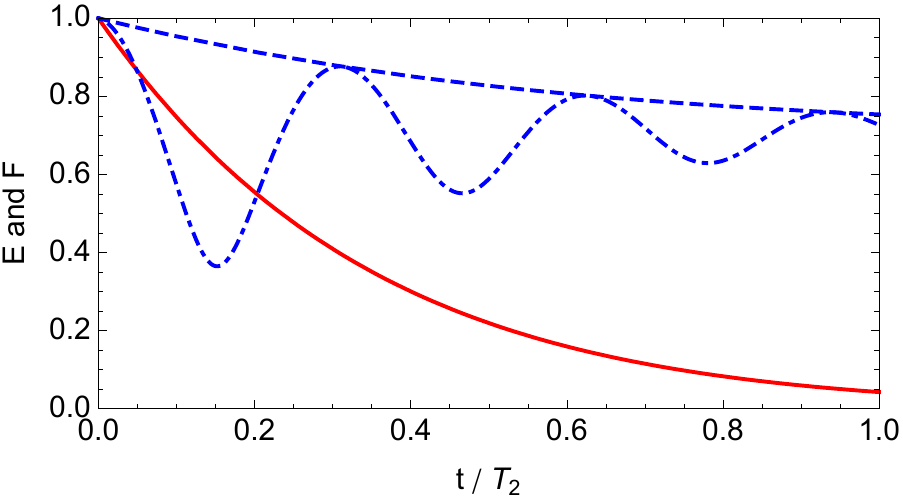}
\caption{Entanglement (solid red) and entanglement fidelity of the $\plus$ (blue dashed) and $\moins$ (blue dot-dashed) 
post-measurement states of the QDs, as a function of the total 
time $t=t_1+t_2+t_3$ of pure dephasing. For the entanglement fidelity of the $\moins$ state we have set $\epsilon=10/T_2$.}
\label{fig:e_and_f}
\end{center}
\end{figure}

In Fig.~({\ref{fig:e_and_f}}) we plot entanglement (red solid curve) and the entanglement fidelities 
$F_{\plus}$ (blue dashed) and $F_{\moins}$ (blue dot-dashed) as a function 
of $t=t_1+t_2+t_3$ in units of the pure-dephasing timescale $T_2$. 
In order for greater levels of entanglement to be reached, all sequence operation times must be kept as small as possible: 
we require that $t_1,t_2,t_3\leq T_2$. 
Without the use of spin echo techniques, $T_2$ times can be as short as ns, primarily due 
to the slowly varying magnetic field induced by the nuclear spins.~\cite{t2_estimate2}
However, spin echo 
techniques are able to extend this timescale up to the $\mu$s range.~\cite{koppens08,t2_estimate} 
As discussed in Ref.~[\onlinecite{bristol_tele}], for QDs in 
micro cavities as we consider the necessary rotations of the QDs can be performed by single photons reflecting 
off the QD-cavity system in the usual way.

\section{Tomography of QD state}

\begin{figure}
\begin{center}
\includegraphics[width=0.325\textwidth]{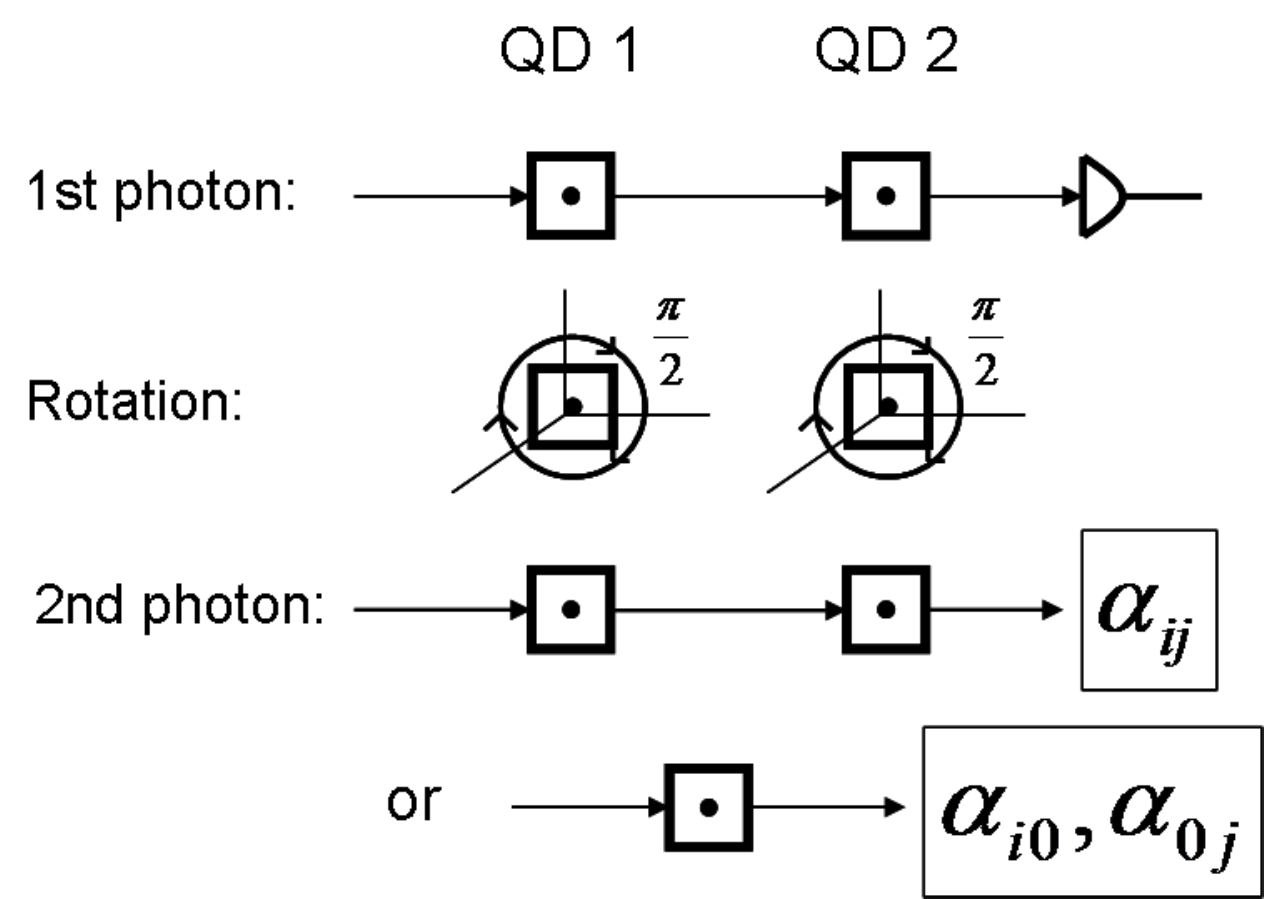}
\caption{Summary of method for QD tomography. The first photon interacts with the QDs and then is measured, leaving the 
QDs in one of two post-measurement states, which we describe by a set of coefficients $\alpha_{ij}$ [see Eq.~({\ref{eq:pauli_expansion}})]. 
We then immediately apply single spin rotations to the QDs. Next, a second photon interacts with one or 
both of the quantum dots. The measurement statistics of the second photon then reveal one of the $\alpha_{ij}$ coefficients, 
depending on which rotations were performed on the QDs.} 
\label{fig:qds_tomo}
\end{center}
\end{figure}

We now propose a method to perform tomography of the post-measurement state of the QDs. 
The general idea is to apply only \emph{single} qubit rotations to the QDs, to then inject a second photon into the 
system, and to determine the state of the QDs through measurement of the second photon. The advantage 
of using only single spin rotations being that they are more easily achieved experimentally, and that they 
cannot affect the level of entanglement shared between the QDs. The tomographic scheme 
is depicted in Fig.~({\ref{fig:qds_tomo}}). 

Recent experiments have 
demonstrated rotations of electrons in QDs on picosecond timescales.~\cite{pico_rot} We therefore assume that the rotations 
we require are effectively instantaneous compared to our other timescales of interest, namely $T_2\sim \mu$s and 
$T_1\sim$ ms. 
We begin by writing a completely general density matrix for the QDs in its Hilbert-Schmidt decomposition: 
\begin{equation}
\rho_{\mathrm{QDs}}=\frac{1}{4}\sum_{i,j=0}^{3} \alpha_{ij}\,\sigma_i\otimes\sigma_j,
\label{eq:pauli_expansion}
\end{equation}
where we define $\alpha_{00}=1$ and $\sigma_0=\openone$. We note 
that the real coefficients $\alpha_{ij}=\mathrm{Tr}(\rho_{QDs} \sigma_i\otimes\sigma_j)$ are expectation values of 
measurements made on the QDs, and are a complete representation of the state.  
Eq.~({\ref{eq:pauli_expansion}}) represents the post-measurement state of the QDs, i.e. 
$\rho_{\Phi}$ or $\rho_{\Psi}$. 
Following the measurement, we immediately apply single spin rotations to the QDs with general 
unitary transformations of the form $R_i(\vec{\theta})=\exp[\frac{i}{2}\vec{\theta}\cdot\vec{\sigma}_i]$, where 
$\vec{\sigma}_i=(\sigma^x_i,\sigma^y_i,\sigma^z_i)$, 
and $\vec{\theta}=(\theta_x,\theta_y,\theta_z)$ is a vector of the rotation angles, and the index refers to the ith QD. 
We write the state of the QDs after the rotations have been performed as 
$\tilde{\rho}_{\rm{QDs}}=R_1(\vec{\theta}_1)R_2(\vec{\theta}_2)\rho_{\rm{QDs}}
R_2(\vec{\theta}_2)^{\dagger}R_1(\vec{\theta}_1)^{\dagger}$, where $\vec{\theta}_1$ and $\vec{\theta}_2$ 
describe the rotations performed on the two QDs. We note that the rotations applied to the QDs are 
equivalent to a transformation of the matrix with elements $\alpha_{ij}$. That is, the rotated state $\tilde{\rho}_{\mathrm{QDs}}$ 
also has a Hilbert-Schmidt decomposition of the same form as Eq.~({\ref{eq:pauli_expansion}}) but with 
coefficients $\tilde{\alpha}_{ij}=\mathrm{Tr}(\tilde{\rho}_{\rm{QDs}}\sigma_i\otimes\sigma_j)$ which are 
functions of the original coefficients $\alpha_{ij}$.

The second photon injected into the system is also vertically polarised, giving 
a complete density operator $\rho_{\rm{tot}}=\ketbra{V}{V}\otimes \tilde{\rho}_{\rm{QDs}}$ before interaction with the QDs. Once 
the second photon has been allowed to interact with the QDs, we trace out the QD degrees of freedom to obtain the reduced 
density operator of the second photon only. 
This reduced density operator is given by,
\beq
\rho_{\gamma}^{(1,2)}=\mathrm{Tr}_{\rm{QDs}}\Big(\e^{\mathcal{L}\tau_3}\big[U_2 \big(\e^{\mathcal{L}\tau_2}\big[U_1\big( \e^{\mathcal{L} \tau_1}
\rho_{\rm{tot}}\big)U_1^{\dagger}\big]\big)U_2^{\dagger}\big]\Big)
\label{rho12}
\eeq
where $\mathrm{Tr}_{\mathrm{QDs}}$ is a trace over the degrees of freedom of the QDs, and the times $\tau_n$ for $n=1,2,3$ parameterise 
the second photon's path through the QD system. The $(1,2)$ superscript indicates that the photon interacts with both QDs. In order to evaluate 
Eq.~({\ref{rho12}}) it is useful to define a projection operator, $\mathcal{P}$, which projects onto the diagonal QD-subspace. i.e. 
we have $\mathcal{P}\rho=\sum_{a}\ketbra{a}{a}\rho\ketbra{a}{a}$ where 
$a\in\{\ket{\uparrow\uparrow},\ket{\uparrow\downarrow},\ket{\downarrow\uparrow},\ket{\downarrow\downarrow}\}$. For 
pure-dephasing noise we find that $\mathcal{P}\mathcal{L}\rho=\mathcal{L}\mathcal{P}\rho=0$, while the form 
of $U_i(\varphi)$ means that $\mathcal{P} U_i(\varphi)\rho U_i(\varphi)^{\dagger}=U_i(\varphi)\mathcal{P} \rho U_i(\varphi)^{\dagger}$. 
Lastly, since $\mathrm{Tr}_{\rm{QDs}}(\dots)=\mathrm{Tr}_{\rm{QDs}}(\mathcal{P}\dots)$, we find that 
$\rho_{\gamma}^{(1,2)}=\mathrm{Tr}_{\rm{QDs}}(U_2 U_1 \mathcal{P}\rho_{\mathrm{tot}}U_1^{\dagger}U_2^{\dagger})$. Using this, 
together with the form of $U_i$ we find the relevant part of the QD-QD-photon state is,
%
%\begin{align}
%U_2U_1 \mathcal{P}\rho_{\mathrm{tot}}&U_1^{\dagger}U_2^{\dagger}
%=\frac{1}{8}\sum_{i,j \in \{0,3\}}\tilde{\alpha}_{ij}\nonumber\\
%&\times\big(\openone^{\gamma}\otimes\sigma_i\otimes \sigma_j-\sigma_x^{\gamma}\otimes\sigma_j\otimes\sigma_i \big),
%\label{TransformedState}
%\end{align}
\begin{align}
U_2U_1 \mathcal{P}\rho_{\mathrm{tot}}U_1^{\dagger}U_2^{\dagger}
=&\textstyle{\frac{1}{8}}\big[\big(\openone^{\gamma}\otimes\openone\otimes \openone-\sigma_x^{\gamma}\otimes\sigma_z\otimes\sigma_z \big)\nonumber\\
+\tilde{\alpha}_{z0}&\big(\openone^{\gamma}\otimes\sigma_z\otimes \openone-\sigma_x^{\gamma}\otimes\openone\otimes\sigma_z \big)\nonumber\\
+\tilde{\alpha}_{0z}&\big(\openone^{\gamma}\otimes\openone\otimes \sigma_z-\sigma_x^{\gamma}\otimes\sigma_z\otimes\openone \big)\nonumber\\
+\tilde{\alpha}_{zz}&\big(\openone^{\gamma}\otimes\sigma_z\otimes \sigma_z-\sigma_x^{\gamma}\otimes\openone\otimes\openone\big)\big]
\label{TransformedState}
\end{align}
where $\openone^{\gamma}=\ketbra{R}{R}+\ketbra{L}{L}$ and 
$\sigma_x^{\gamma}=\ketbra{R}{L}+\ketbra{L}{R}$ act on the photon, and the coefficients 
$\tilde{\alpha}_{ij}=\mathrm{Tr}(\tilde{\rho}_{\rm{QDs}}\sigma_i\otimes\sigma_j)$ pertain to the rotated QD-QD density operator. 
Taking the trace of Eq.~({\ref{TransformedState}}) gives 
\begin{equation}
\rho_{\gamma}^{(1,2)}=\frac{1}{2}\left(\begin{array}{cc}
1&-\tilde{\alpha}_{zz}\\
-\tilde{\alpha}_{zz}&1
\end{array}\right),
\label{eq:photon_state}
\end{equation}
where the matrix is written in the circularly polarised $\{|R\rangle,|L\rangle\}$ basis. 
%The real quantity $\xi$ is in general a function of the set of 
%coefficients $\alpha_{ij}$, the rotation angles $\vec{\theta}_1$ and $\vec{\theta}_2$, and the times $\tau_j$.  
Eq.~({\ref{eq:photon_state}}) is the reduced density operator of the second photon following its interaction with the two 
QDs. Remarkably, we see  
that for the pure-dephasing noise considered here, this state does not depend on any of the time intervals $\tau_n$ 
describing its path through the QD system. Thus, any pure-dephasing noise affecting the QDs after the necessary 
rotations have been applied, does not affect the accuracy of the tomographic scheme. This is a consequence 
of the properties of the projection operator $\mathcal{P}$ used above, that 
$U_1$ and $U_2$ couple only to the diagonal elements of $\tilde{\rho}_{\rm{QDs}}$, and that the 
pure-dephasing noise does not affect these elements. 
%This can be appreciated 
%by noting that $U_1$ and $U_2$ couple only to the diagonal elements of $\chi_{\rm{QDs}}$, which are 
%unaffected by pure-dephasing noise. As such, the trace in Eq.~({\ref{rho12}}) picks out density matrix elements which 
%are protected from noise.

Measurement of the second photon in the usual linearly polarised basis, $\{\ket{V},\ket{H}\}$, leads to outcome probabilities of the 
form $P_\plus=\frac{1}{2}(1-\tilde{\alpha}_{zz})$ and $P_\moins=\frac{1}{2}(1+\tilde{\alpha}_{zz})$ for the $\plus$ and $\moins$ outcomes, respectively. 
Analysing the statistics of this measurement it is therefore possible to extract the value of $\tilde{\alpha}_{zz}=P_\moins-P_\plus$. 
Since the rotations applied to the QD-QD state before the injection of the second photon 
amount to a rotation of the matrix with elements $\alpha_{ij}$ into that with elements $\tilde{\alpha}_{ij}$, we see 
that the second photon ultimately carries information regarding the original unrotated state $\rho_{\mathrm{QDs}}$. Specifically, 
with the rotations $X=\exp(-\frac{i}{2}\frac{\pi}{2}\sigma_x)$ and $Y=\exp(\frac{i}{2}\frac{\pi}{2}\sigma_y)$ (and the 
identity) on dots $1$ and $2$, we find that $\tilde{\alpha}_{zz}\to\alpha_{ij}$ with $i,j\neq0$, see Tab.~(\ref{tab:rot_alpha1}). Thus, 
we see that single spin rotations and a second photon can be used to probe the state of the QDs.

We also note that the procedure described gives information regarding correlations in the QD-QD state. For example, 
when no rotations are performed on the QDs, we have $\tilde{\alpha}_{zz}=\alpha_{zz}$. This is perhaps to be expected, since 
from Eq.~({\ref{eq:unitary}}) we see that the QD-photon interaction depends on the spin projection along the z-axis. Thus, 
we expect a photon having interacted with both QDs to depend on the correlation of the spin-projections, $\alpha_{zz}$. When rotations 
are performed an element $\alpha_{ij}$ with $i,j\neq0$ is moved into the the position $\tilde{\alpha}_{zz}$ leading to information regarding a different correlation. 
For example, with the rotation $Y$ applied to both QDs, we find $\tilde{\alpha}_{zz}=\alpha_{xx}$.

\begin{table}
\begin{center}
\caption{Those $\alpha_{ij}$ coefficients that can be obtained from the reduced density matrix of the second photon by applying  
rotation(s) to the QDs.}
\label{tab:rot_alpha1}
\begin{tabular}{cccc}
\hline
QD$_1$&QD$_2$&\quad&$\tilde{\alpha}_{zz}$\\
\hline
$\mathbb{1}$&$\mathbb{1}$&$\longrightarrow$&$\alpha_{zz}$\\
$\mathbb{1}$&X&$\longrightarrow$&$\alpha_{zy}$\\
$\mathbb{1}$&Y&$\longrightarrow$&$\alpha_{zx}$\\
X&$\mathbb{1}$&$\longrightarrow$&$\alpha_{yz}$\\
Y&$\mathbb{1}$&$\longrightarrow$&$\alpha_{xz}$\\
X&Y&$\longrightarrow$&$\alpha_{yx}$\\
Y&X&$\longrightarrow$&$\alpha_{xy}$\\
X&X&$\longrightarrow$&$\alpha_{yy}$\\
Y&Y&$\longrightarrow$&$\alpha_{xx}$\\
\hline
\end{tabular}
\end{center}
\end{table}

%From Tab.~({\ref{tab:rot_alpha1}}) we see that applying no rotations to the QDs results in the second photon containing 
%information about the expectation value $\alpha_{zz}=\mathrm{Tr}(\rho_{\rm{QDs}}\sigma_z\otimes\sigma_z)$. This can be attributed to the 
%form of the QD-photon interaction, Eq.~({\ref{eq:unitary}}). The QD-photon interaction 
%depends on the spin projection of the QD along the z-axis. Thus, a photon having interacted with two QDs in such a way, will 
%have a state whose form depends on the correlation of the spin projections, $\alpha_{zz}$. 
%Rotations applied to the QDs amount to a rotation of the matrix with elements $\alpha_{ij}$. The rotations 
%therefore move a new element into the position which was previously $\alpha_{zz}$. It is then this new element 
%which is protected from pure-dephasing noise, and  
%this element which the photon state then depends on. 

In order to obtain the remaining $\alpha_{i0}$ and $\alpha_{0j}$ coefficients (the Bloch vector elements of each QD), 
we allow the second photon sent through to 
interact with only one of the two QDs. In place of Eq.~({\ref{rho12}}) we then have, 
\beq
\rho_{\gamma}^{(1)}=\mathrm{Tr}_{\rm{QDs}}\Big(\e^{\mathcal{L}\tau_3}\big[ \e^{\mathcal{L}\tau_2}\big[U_1\big( \e^{\mathcal{L} \tau_1}
\rho_{\rm{tot}}\big)U_1^{\dagger}\big]\big]\Big)
\label{rho1}
\eeq
with a similar expression for $\rho_{\gamma}^{(2)}$. Using methods similar to those used above, 
we obtain a reduced density matrix for the second photon of the form, 
\begin{equation}
\rho_{\gamma}^{(1)}=\frac{1}{2}\left(\begin{array}{cc}
1&i\tilde{\alpha}_{z0}\\
-i\tilde{\alpha}_{z0}&1
\end{array}\right),
\label{eq:photon_state2}
\end{equation}
with an similar expression for $\rho_{\gamma}^{(2)}$ but with $\tilde{\alpha}_{z0}$ replaced with $\tilde{\alpha}_{0z}$. 
Once again, since the diagonal elements of Eq.~({\ref{rho1}}) which the trace picks out are unaffected by pure-dephasing noise, 
we see that the reduced photon state does not depend on $\tau_n$. 
Choosing combinations of the rotations $X$ and $Y$, we find that $\tilde{\alpha}_{z0}$ (or $\tilde{\alpha}_{z0}$) 
can be made equal to the remaining pre-rotation coefficients of the QD-QD state [see Tab.~(\ref{tab:rot_alpha2})]. 
To extract the value of $\tilde{\alpha}_{z0}$ from the 
photon, we now need to measure the photon in the diagonal $|+45^\text{o}\rangle=(|R\rangle+i|L\rangle)/\sqrt{2}$ 
and $|-45^\text{o}\rangle=(|R\rangle-i|L\rangle)/\sqrt{2}$ basis. 
The outcome probabilities would then be $P_{+45^\text{o}}=\frac{1}{2}(1-\tilde{\alpha}_{z0})$ and 
$P_{-45^\text{o}}=\frac{1}{2}(1+\tilde{\alpha}_{z0})$, and $\tilde{\alpha}_{z0}$ can 
be found from the quantity $P_{-45^\text{o}}-P_{+45^\text{o}}$. 

An extremely useful property of our tomographic procedure is that any further photons injected into the 
system after the second will be measured having the same polarisation as the second with unit probability. To see this, 
we consider the state of the QDs after we have injected and measured a second photon in order to measure a QD-QD correlation 
$\alpha_{ij}$ with $i,j\neq 0$. The relevant (diagonal) part of the QDs following measurement of the second photon is  
$\mathcal{P}\chi=(1/p_k)\mathrm{Tr}_{\mathrm{ph}}(\pi_k U_2U_1 \mathcal{P}\rho_{\mathrm{tot}}U_1^{\dagger}U_2^{\dagger})$, for 
$k=\Phi,\Psi$. Using Eq.~({\ref{TransformedState}}) we can see that this state is one 
having $\mathrm{Tr}(\chi \sigma_z\otimes\sigma_z )=\alpha_{zz}=1$ or $\alpha_{zz}=-1$, depending on the two measurement outcomes $k=\Phi$ and $k=\Psi$ respectively. 
Therefore, if an additional photon is injected (after the second) without performing any rotations of the QDs, since 
$P_\plus=\frac{1}{2}(1-\alpha_{zz})$ and $P_\moins=\frac{1}{2}(1+\alpha_{zz})$, this additional photon will 
be measured having the same polarisation as the second with unit probability. We note that since the measurement probabilities depend 
only on $\alpha_{zz}$ this result is independent of the precise form of the post-measurement QD-QD state. The process can 
be repeated continuously to 
build up a string of photons all of which will be measured having the same polarisation, with the probability 
that this polarisation is $H$ or $V$ corresponding precisely to the value $\alpha_{ij}$ determined by the rotations 
performed after the first entangling photon. A similar property is also true when we measure 
elements $\alpha_{i0}$ and $\alpha_{0j}$, where now the string of photons will all have polarisation $\ket{-45^\text{o}}$ or $\ket{+45^\text{o}}$.

\begin{table}
\begin{center}
\caption{Coefficients $\alpha_{0i}$ and $\alpha_{i0}$ that can be obtained from the reduced density matrix of the second photon by applying a 
rotation to one QD before the second photon interacts only with one of the QDs.}
\label{tab:rot_alpha2}
\begin{tabular}{cccc}
\hline
QD$_1$&QD$_2$&\quad&$\tilde{\alpha}_{z0}$  or $\tilde{\alpha}_{oz}$\\
\hline
$\mathbb{1}$&(no interaction)&$\longrightarrow$&$\alpha_{z0}$\\
X&(no interaction)&$\longrightarrow$&$\alpha_{y0}$\\
Y&(no interaction)&$\longrightarrow$&$\alpha_{x0}$\\
(no interaction)&$\mathbb{1}$&$\longrightarrow$&$\alpha_{0z}$\\
(no interaction)&X&$\longrightarrow$&$\alpha_{0y}$\\
(no interaction)&Y&$\longrightarrow$&$\alpha_{0x}$\\
\hline
\end{tabular}
\end{center}
\end{table}

Thus, we see that with only single spin rotations of the QDs, we are able to reconstruct the complete post-measurement state. Moreover, 
owing to the form of the QD-photon interaction, pure-dephasing noise affecting the QDs after the rotations have been preformed has no bearing on the 
accuracy of the tomographic procedure. Thus, while it is important for the initial photon's path through the QD system to be achieved 
on a timescale shorter than $T_2$ (in order to generate an appreciable amount of entanglement), the amount of 
entanglement that is measured by the second photon reflects the true amount that was present immediately after the rotations are performed. 
Additionally, any photons injected after measurement of the second will be measured 
having the same polarisation as the second with unit probability, and can therefore be used to strengthen the signal. 

It should be noted that the arguments above hold only for pure-dephasing $T_2$-type noise: the details of the tomographic process 
mean that the $T_2$ timescale is unimportant. As such, it is unnecessary to employ any spin echo techniques to lengthen $T_2$, 
since the important timescale becomes the spin relaxation timescale $T_1$. Unlike pure-dephasing processes, spin relaxation 
processes do affect the diagonal elements of the QD-QD density matrix. As such, if the parameters describing the 
second photon's path through the QD system, $\tau_n$, were 
of the order of the spin relaxation timescale, $T_1$, we would find that the reduced state of the second photon does depend on $\tau_n$. 
Thus, for accurate tomography to be achieved, it is still necessary that the second photon traverses the QD system on a timescale $\ll T_1$.

\section{Spin-relaxation}

Though it seems that with current technologies the dephasing timescale of electron spins in QDs is likely to be far shorter than the spin-relaxation timescale, 
it is nevertheless interesting to briefly investigate what differences may occur if the coherence of the QDs were limited by relaxation processes. 
In order to do so, we now consider the set of Lindblad operators given by 
$L_i=\sqrt{\Gamma_1}\sigma^-_i$ for $i=1,2$, 
where $\sigma^-_i=|{\downarrow}\rangle\langle{\uparrow}|_i$ 
and $T_1^{-1}=\Gamma_1$ is the spin-relaxation time.

Our current protocol consists of the injection and measurement of a first photon to establish entanglement, the application of rotations to the 
QDs, followed by the injection and measurement of a string of photons whose polarisations reveal information about the QD state established. 
For noise originating from spin relaxation processes, 
the first stage of this protocol is largely unaffected. For appreciable levels of entanglement to be achieved, the first photon's 
path through the QD system, characterised by the times $t_n$, must be short, but now compared to the spin-relaxation time $T_1$. We do note, 
however, that now the measurement outcome probabilities do change, and we have 
\begin{equation}
P_\plus=\frac{1}{2}\e^{-t_1\Gamma_1}\left(1+\e^{-t_2\Gamma_1}-\e^{-(t_1+t_2)\Gamma_1}\right),
\label{eq:prob_plus}
\end{equation}
and $P_\moins=1-P_\plus$. 
Thus, in this case we see that the three time intervals play a decreasingly important role in the evolution of the probabilities. We also see 
that as $t_1$ and $t_2$ increase the probability of obtaining the $\moins$ outcome tends to one while that of the $\plus$ outcome tends to 
unity. This can be understood by noting that spin relaxation transfers population to the state $\ket{\downarrow\downarrow}$ which has 
overlap only with $\ket{\Phi}$.

For spin-relaxation we find that the tomographic procedure described in the previous section still works, though does not share some 
of the helpful features previously described. Namely, while a reduced photon state of the form of Eq.~({\ref{eq:photon_state}}) is found, 
for the combinations of rotations described in Tab.~({\ref{tab:rot_alpha1}}), the value 
$\tilde{\alpha}_{zz}$ becomes a combination of the original $\alpha_{ij}$ values, as well as the times describing the second photon's path through the system. 
We note, however, that tomography ought still to be possible, but will require specific knowledge of $T_1$, and the ability to vary 
the time taken for the second photon to pass through the system.

We lastly note an intriguing feature of spin-relaxation noise, which is that it allows 
for the level of entanglement to be boosted by additional photons which pass through the system and measured. Numerical simulations 
suggest that if photons can be sent through the system and measured at a sufficient rate, sequences of measurement outcomes 
all being $\plus$ will correspond to a state of the QDs with entanglement maintained at a particular level. We believe future research 
in this direction to be worthwhile, though beyond the scope of this work.

\section{Summary}

In Ref.~[\onlinecite{bristol_dots}] a scheme to entangle two spatially separated QDs in micro cavities was proposed. The scheme 
relies on a QD-state dependent rotation of the polarisation of a photon which is exchanged. In this work we have investigated the 
effects of non-unitary dynamical evolution of the QDs, caused by the coupling to their solid-state environment. We found that  
pure-dephasing of the QDs necessarily decreases the level of entanglement that can be attained.

We then proposed a method to perform tomography of the state of the QDs, which relies on only single spin local rotations of the QDs, 
and the injection of additional photons. Interestingly, while the level of entanglement attained is sensitive to pure-dephasing noise, 
the accuracy of the tomographic procedure is not. Thus, the time in which to perform the tomography is limited only by 
the spin relaxation timescale, which is typically orders of magnitude greater than the pure-dephasing time. Lastly, we 
found that within the pure-dephasing time, many photons can be injected into the system in order to boost the tomographic measurement 
signal.

\section{Acknowledgements}

The authors wish to thank Cl\'{e}ment Stahl for many useful discussions. D.P.S.M. thanks CONICET, and together with 
T.R. also acknowledges support from the 
EPSRC and CHIST-ERA project SSQN.

%\bibliography{refs}

\end{document}